\pdfoutput=1
\documentclass[longauth]{aa} 

\usepackage{natbib}
\usepackage{graphicx}
\usepackage{txfonts}
\usepackage{amstext}

\begin{document}

   \title{Exocomet signatures around the  A-shell star $\phi$ Leo?}


   \author{C. Eiroa
          \inst{1}
\and
          I. Rebollido\inst{1}
\and
          B. Montesinos\inst{2}
\and
           E. Villaver\inst{1}
\and       O. Absil\inst{3}
\and       Th. Henning\inst{4}
\and       A. Bayo\inst{5}
\and       H. Canovas\inst{1}
\and       A. Carmona\inst{6}
\and       Ch. Chen\inst{7}
\and       S. Ertel\inst{8}
\and       D. P. Iglesias\inst{5}
\and       R. Launhardt\inst{4}
\and       J. Maldonado\inst{9}
\and       G. Meeus\inst{1}
\and       A. Mo\'or\inst{10}
\and       A. Mora\inst{11}
\and       A. J. Mustill\inst{12}
\and       J. Olofsson\inst{5}
\and       P. Riviere-Marichalar\inst{13}
\and       A. Roberge\inst{14}
          }
   \institute{Dpto. F\'\i sica Te\'orica, Universidad Aut\'onoma de Madrid, 
Cantoblanco,
              28049 Madrid, Spain,
              \email{carlos.eiroa@uam.es}
\and CAB (CSIC-INTA), Camino Bajo del Castillo, s/n, 28692 Villanueva de la 
Ca\~nada, Madrid, Spain
\and STAR Institute, Universit\'e de Li\`ege, F.R.S.-FNRS, 19c All\'ee du Six Ao\^ut, B-4000 Li\`ege, Belgium
\and Max-Planck-Institut für Astronomie (MPIA), Königstuhl 17, D-69117 Heidelberg, Germany
\and Instituto de Física y Astronomía, Facultad de Ciencias, Universidad de 
Valparaíso,  5030 Casilla, Valparaíso, Chile
\and Universit\'e de Toulouse, UPS-OMP, IRAP, Toulouse F-31400, France
\and Space Telescope Science Institute, 3700 San Martin Drive, Baltimore, MD 21212, USA
\and Steward Observatory, Department of Astronomy, University
of Arizona,  Tucson, AZ 85721, USA
\and INAF, Osservatorio Astronomico di Palermo, 90134 Palermo, Italy
\and Konkoly Observatory, Research Centre for Astronomy and Earth Sciences, 
PO Box 67, 1525 Budapest, Hungary
\and Aurora Technology B.V. for ESA, ESA-ESAC,
Camino Bajo del Castillo, s/n, 28692 Villanueva de la
Ca\~nada, Madrid, Spain
\and Lund Observatory, Department of Astronomy and Theoretical Physics, Lund
University, Box 43, SE-221 00 Lund, Sweden
\and ESA-ESAC, Camino Bajo del Castillo, s/n, 28692 Villanueva de la
Ca\~nada, Madrid, Spain
\and Exoplanets \& Stellar Astrophysics Lab, NASA Goddard Space Flight Center, 
Code 667, Greenbelt, MD 20771, USA 
             }

   \date{}

 
  \abstract
{We present an intensive monitoring  of high-resolution spectra of the
  Ca {\sc ii} K  line in the A7IV shell star $\phi$  Leo at very short
  (minutes, hours),  short (night to  night), and medium  (weeks,
  months)   timescales.    The   spectra  show   remarkable   variable
  absorptions  on   timescales  of   hours,  days,  and   months.   The
  characteristics of these sporadic events are very similar to most that are observed toward  the debris disk  host star  $\beta$ Pic,
  which are commonly interpreted as  signs of the evaporation of solid,
  comet-like bodies grazing or falling  onto the star.  Therefore, our
  results  suggest the  presence of  solid bodies  around $\phi$  Leo.
  To  our knowledge, with the exception of
  $\beta$ Pic, our monitoring has the best time resolution
  at the mentioned timescales for a star with events attributed to
  exocomets. Assuming  the cometary  scenario and considering the timescales 
of our monitoring, our  results indicate that  $\phi$  Leo
  presents the richest environment with comet-like events known
to date,
  second only to $\beta$ Pic. 
}


   \maketitle
%
\section{Introduction}
In spite of their low mass,  Kuiper Belt objects, comets, and asteroids
are key elements for understanding the  early history of the solar system,
its  dynamics, and  composition.   While exoplanets  are now  routinely
detected and  hundreds of  debris disks  provide indirect  evidence of
planetesimals  around  main-sequence  (MS)  stars  \citep{matthews14},
little is  directly known about  minor bodies around other  stars than
the Sun.  The immense difficulty of  a direct detection lies
in their lack of  a large  surface area, which is  required to  detect their
thermal or scattered  emission. Dust features provide  hints about the
properties  of  $\mu$m-sized  grains  in debris  disks that result  from
collisions of planetesimals \citep[e.g.][]{olofsson12}.  Circumstellar
(CS) CO  emission around some  AF-type MS stars \citep[e.g.,][]{moor15,marino16} has  been interpreted to be the result of outgassing produced
by  comet collisions  \citep{zuckerman12}.   A complementary,  somehow
more direct, information  on the exocomet nature  and composition is
provided  by  the  detection  of variable  absorptions  superposed  on
photospheric lines in the spectra of some stars.

Variable absorption  features in  metallic lines  have been  known for
about   30   years  in   the   optical   spectrum  of   $\beta$   Pic
\citep{hobbs85}.   These  irregular  features, mainly  traced  in  the
Ca~{\sc  ii}  K  line,  appear  redshifted  and to   a  much  lower  degree
blueshifted  with respect  to the  radial velocity  of the  star, and
might vary on timescales as short as hours. These features have been interpreted
to be the result of the gas released  by the evaporation of exocomets grazing or
falling      onto       the      star       \citep[and      references
  therein]{ferlet87,kiefer15} that are driven into the vicinity of the star by
the   perturbing   action   of   a  larger   body,   that is,
by   a   planet
\citep{beust91}. Variable  absorptions  like this have also  been
observed      toward     several A-type      stars
\citep[e.g.,][]{redfield07,roberge08,welsh15}.

We   have   initiated   a  short-   and   medium-term   high-resolution
spectroscopic project aiming at detecting and monitoring these sporadic
events that are attributed to  exocomets in a sample of MS  stars, most of them
A-type stars, but also some  FG-type stars with a range  of ages.  The sample
includes  stars  for which  exocomet  signatures  have  been
detected previously, such as 49 Ceti or HR 10,  and also stars that have not been scrutinized yet
for such events.   So far, we have obtained more  than 1200 spectra of
$\text{about }$100 stars (Rebollido et al., in preparation). The star $\phi$  Leo (HR 4368, HD 98058, HIP  55084) 
stands out  as its
spectrum exhibits conspicuous variations  on timescales of hours, days,
and months.   
This  work presents our results  concerning the Ca {\sc ii} K line obtained 
in five observing runs  from December 2015 to May 2016. 
This line is particularly sensitive to these absorptions and
is most frequently   analyzed in  the exocomet  literature.  
Results
for other relevant lines, including the Ca {\sc ii} IR-triplet, Ti
{\sc ii}, Na {\sc i} D, and  Balmer lines, together with the spectra of
other stars, will be presented in a forthcoming paper.

\section{$\text{}\phi$ Leo: properties and astrophysical parameters}
\label{hr4368}

$\text{The source }\phi$  Leo is  an A7IVn  shell  star \citep{jaschek91}  located at  a
distance of 56.5  pc.  The star is  seen close to edge-on  with a very
high   rotational   velocity,  $v \sin   i$   $\sim$220   -  254   km/s
\citep{lagrangehenri90,royer07}.   It is  surrounded by  a gaseous  CS
disk detected in the  Ti {\sc ii} 3685, 3759, 3762  \AA ~lines and       in      the       Ca      {\sc       ii}      H/K       lines
\citep[e.g.,][]{jaschek88,abt08}. The star $\phi$ Leo is  close to the center of
the local interstellar bubble, a cavity of low density, which makes it
unlikely  that   its  shell or disk  is   formed  by  accretion   of  the
interstellar   medium,  as   conjectured  for   other  A-shell   stars
\citep{abt15}.  The Ca {\sc ii}  K profile shows a triangular shape
probably  due to  the  combination  of the  photospheric  and disk
absorptions \citep{lagrangehenri90}.  The star  does not posses a warm
debris disk  \citep{rieke05}, although a cool debris disk  cannot be 
excluded since data  at $\lambda >$ 25 $\mu$m 
do not exist, to our knowledge.  The best Kurucz photospheric model
fitting our spectra is $T_{\rm eff}$ = 7500  K, $\log g$ = 3.75, $v\sin i$
=  230 km/s,  in  good  agreement with  
previous estimates 
\citep[e.g.,][]{david15}.   Its  bolometric  luminosity  is  $\sim$  45
L$_\odot$ \citep{zorec12}, and age estimates  are in the  range $\sim$
500-900 Myr \citep{david15,derosa14,zorec12}.

\section{Observations}

High-resolution spectra  of $\phi$  Leo were  obtained with  the high-resolution fiber-fed \'echelle spectrographs HERMES and FEROS attached
to the Mercator (La Palma, Spain)  and MPG/ESO 2.2 m (La Silla, Chile)
telescopes, respectively.   A total of  60 spectra of $\phi$  Leo were
taken. Observing dates, telescope, and number of spectra per night are
given in Table  \ref{table:log}.  HERMES has a  spectral resolution of
$\sim$  85000  (high-resolution  mode) covering  the  range  $\lambda
\lambda$ $\sim$370-900 nm ~\citep{hermes}.  Exposure times ranged from
260  to  600  seconds.   Some spectra  were  taken  consecutively  and
combined   after  reduction.    FEROS   has   a  spectral   resolution
R$\sim$48000   covering   the  range   from   $\sim$350   to  929   nm
\citep{feros}.  Exposure times  ranged from 120 to  360 seconds.  The
HERMES  and  FEROS  spectra  were  reduced  using  the  available
pipelines of the two instruments.  Barycentric corrections were also
made for all spectra.

\begin{table}
\caption{Log of observations}
\label{table:log}
\centering
\begin{tabular}{lll}
\hline
Observing run & Dates & Spectra per night      \\
\hline
2015 December$^1$ & 20, 22, 23     & 1, 4, 1   \\
2016 January$^1$  & 27, 28, 30     & 1, 3, 3   \\
2016 March$^1$    & 3, 4, 5, 6 & 4, 4 ,2 ,3 \\
2016 March$^2$    & 25, 26, 27, 28 & 4, 4, 3, 3 \\
2016 May$^1$      & 11             & 20         \\
\hline
\end{tabular}

$^1$ HERMES, $^2$ FEROS
\end{table}

\section{Results}
Distinct changes  are observed  in the Ca  {\sc ii} H  and K  lines of
$\phi$ Leo.   These changes  affect the  line profile  as well  as the
wavelength  and depth  of the  absorption peak.  Variations occur  on
timescales  of  hours, days,  and  months.  All  the spectra  show  the
triangular shape  noted by \cite{lagrangehenri90} , modulated  in most
cases  by additional  redshifted  absorption.  Furthermore, we found no 
temporal  variability pattern.  
In  the following we  refer to the Ca  {\sc ii} K line  alone, although we also comment  
on some Ca  {\sc ii}  H line spectra 
to illustrate  the results
for this line in comparison with the K line.

\subsection{Stable  Ca {\sc ii} K component} 

\begin{figure} 
\centering
\scalebox{0.33}{\includegraphics[angle=-90]{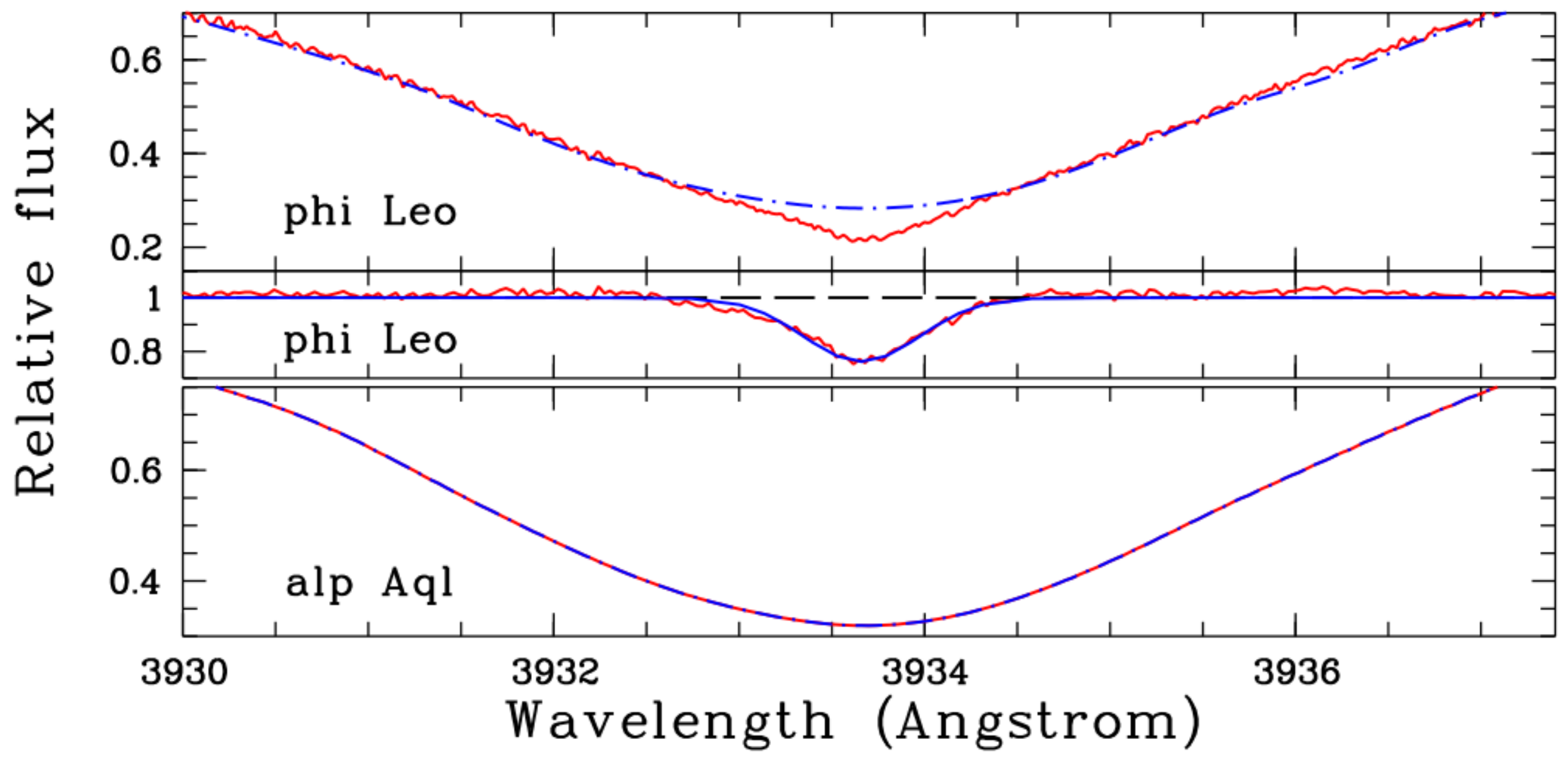}}
\caption{Top:  Average Ca  {\sc ii}  K line  of $\phi$  Leo from  2015
  December 20  and 23  (red continuous line),  together with  a Kurucz
  photospheric  model  (blue dash-dotted  line).   See  text for  stellar
  parameters. Middle: residuals of the  average 20 and 23 spectra with
  respect  to the  photospheric line.  The  blue continuum  line is  a
  Gaussian with  FWHM = 56 km/s.  Bottom: observed Ca {\sc  ii} K line
  (red  line) of  $\alpha$  Aql  compared with  a  Kurucz model  (blue
  dash-dotted line) with stellar parameters  $T_{\rm eff}$ = 7900 K, $\log g$
  = 4.25, $v\sin i$ = 210 km/s. }
\label{hr4368_k} 
\end{figure}

\begin{figure*} 
\centering
\scalebox{0.33}{\includegraphics[angle=-90]{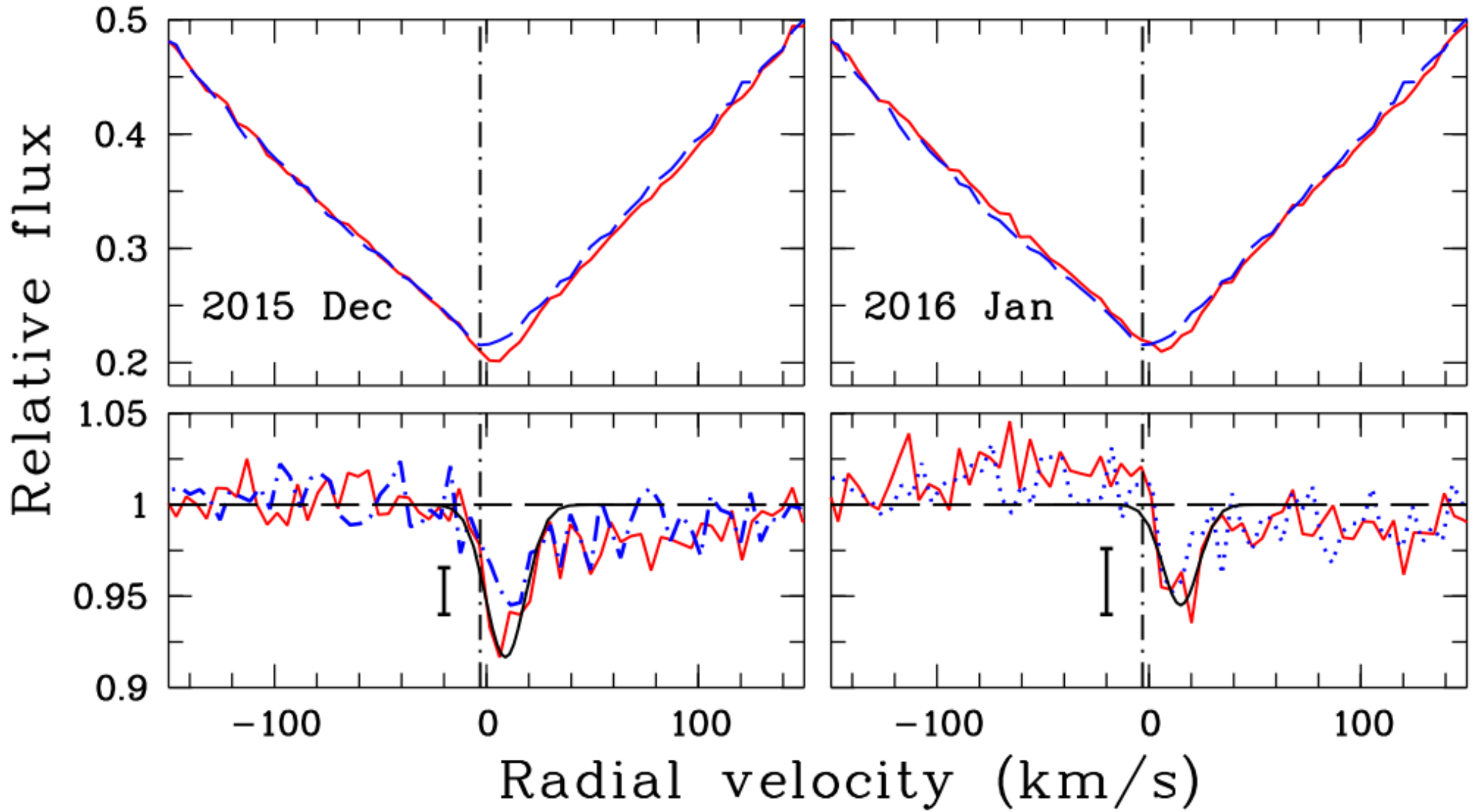}}
\scalebox{0.34}{\includegraphics[angle=-90]{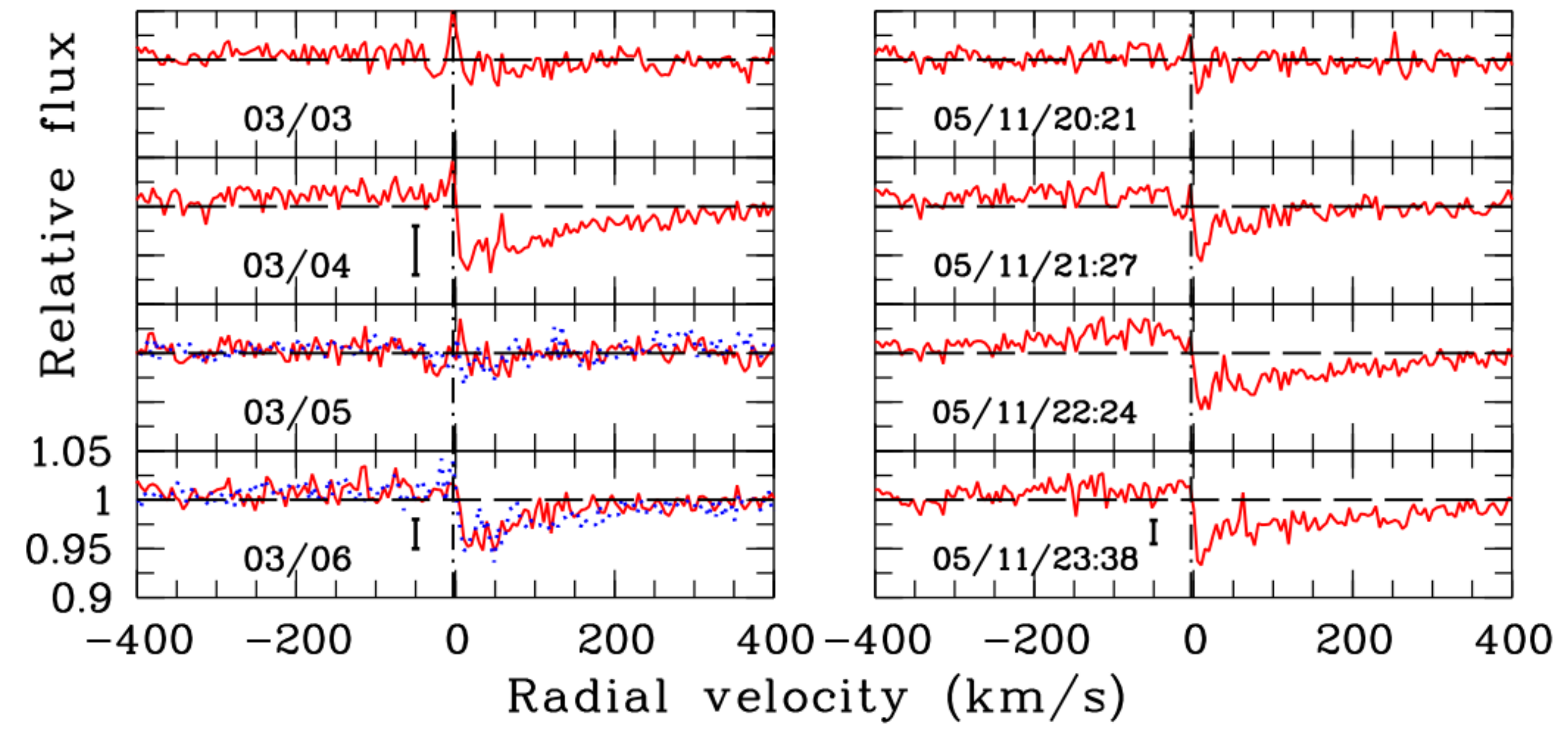}}
\caption{From left to right the Ca  {\sc ii} K line from the observing
  runs with  Mercator. At the top  of the December and  January panels
  the average of  the four December 22 spectra and  the average of the
  January spectra are plotted (red continuous lines) together with the
  template  spectrum  (blue   dashed  line).  At  the   bottom   the
  corresponding residuals are shown in red. To  guide the eye, two  Gaussians with FWHM
  25 km/s (December)  and 20 km/s (January) are plotted  (black continuous 
line).  The
  March  panel shows  the residuals  of the  average spectrum  of each
  night. The May  panel shows the average of five  consecutive spectra of
  May 11 starting at the UT  indicated in the panel.  Some panels show
  the residuals of  the Ca {\sc ii} H line  for comparison (blue dotted line).
  A 3-$\sigma$ error bar  is plotted in some panels. The dash-dotted line at
  -3.0 km/s corresponds to the radial velocity of the star.}
\label{mercator} 
\end{figure*}
The Ca {\sc  ii} K line spectra  taken on 
2015 December 20
and 23 are  identical with their absorption peaks at  the stellar 
radial velocity,
 $v_{\rm rad}$ = -3 km/s \citep{bruijne12}.  The  remaining
spectra are either  deeper, with the absorption peak  shifted to longer
wavelengths, or  with obvious additional redshifted  absorption components
superposed on the December 20  and 23 spectra.  The exceptions are
the spectra of 2016 March 3 and 5, which show small differences with those of
December 20 and 23 (see below).

Figure \ref{hr4368_k}  shows the average  of the  December 20 and  23 Ca
{\sc ii}  K line together with the  photospheric line of  the Kurucz  model with
stellar parameters mentioned in  Sect. \ref{hr4368}. The broad
additional absorption with a  FWHM
=  56  km/s   at  approximately the  stellar  radial
velocity produces the triangular shape of the line profile. Its depth and 
equivalent width are 0.238 and  0.22  \AA, respectively. This corresponds to a 
column density $N({\rm Ca}$ {\sc ii}) $\sim$$2.3\times10^{12} ~{\rm cm}^{-2}$. 
To  ensure that  this additional absorption  is real,  we  synthesized a photospheric model for $\alpha$ Aql, a star with similar
spectral type (A7Vn) and rotational velocity,  and compared it to spectra taken  with the same
instrument and  configuration as  for $\phi$ Leo.  Figure \ref{hr4368_k}
shows the excellent fit of the corresponding $\alpha$ Aql model to the
observed spectra.

These results suggest that during December  20 and 23 we detected
the  contribution  of  a  CS   gas  disk  superposed  on  the  stellar
photosphere, while the  remaining spectra are  affected by additional
gas  absorptions,  although  other  causes  producing  the  triangular
absorption cannot be excluded (see  below). In the following we assume
that the additional  absorption seen on December 20 and  23 is a stable
component  in addition  to the  $\phi$  Leo photosphere  and take  the
average of the two  spectra as a template to  analyze other contributions
to the gas components.

\subsection{Variable  Ca {\sc ii} K components} 
  
{\em Mercator  2015 December.}  The four  spectra of December 22  show no
appreciable changes, but their absorption  is redshifted and deeper than
the average
December   20    and   23    template   spectrum.
Figure \ref{mercator}  shows the average of  these four spectra and 
the residuals, taking the template  spectrum as a continuum for the
spectrum of December 22. An  obvious absorption event occurs
at v $\sim$6.0
km/s,  that is,  $\sim$9.0  km/s redshifted  from  the  stellar  radial
velocity, and depth $\sim$0.083. A second  event might be present at a
velocity $\sim$16 km/s and depth  $\sim$0.060. The equivalent width of
both events  together is   $\sim$23 m\AA,  and its FWHM is $\sim$25
km/s. The figure also shows the residuals corresponding to the Ca {\sc
  ii} H; in  this case, only one  event with a depth  of $\sim$0.054 is
clearly seen.

\noindent {\em  Mercator 2016  January.}  These  spectra appear  slightly
redshifted with respect to the  template spectrum, but  their depth  is lower
than the  one of December 22.   There might be tiny  changes among the
spectra  taken on  different nights,  although we  consider only  their
average because  poor weather  conditions produced relatively  low signal-to-noise (S/N)
ratios.  Figure \ref{mercator}  shows the 2016 January  spectrum and its
residual  compared to  the  template.   An event  at  $\sim$15  km/s,
that is, redshifted  by $\sim$18 km/s  with respect to the stellar radial  velocity, is
clearly discernible.  Its equivalent width is 20 m\AA, the depth
is 0.055, and the
FWHM is $\sim$20 km/s.

\noindent {\em Mercator 2016 March.} No detectable variations
were recorded during the individual nights, but the Ca {\sc ii} K line clearly varied
from   night  to   night.  Figure   \ref{mercator}  shows   the  nightly
residuals. 
The  spectra of March 3 and 5 are
similar to  the template, but a  clear flux  depression is evident
on
March 4 and  6 at $\sim$16  km/s, the   March 4
depth of $\sim$0.065 is deeper than the depth on March 6 of $\sim$0.047. In both
cases,  a  broad  wing  extending  up to  $\sim$200  km/s  might  be
present. The  corresponding Ca {\sc  ii} H  line depths are  0.052 and
0.047 for March 4 and 6, respectively.

\noindent {\em Mercator 2016 May.} Twenty spectra were taken consecutively
during four hours  on 2016 May 11. Figure  \ref{mercator}  shows the Ca {\sc
  ii} K residuals  with respect to the template
spectrum grouped in four one-hour periods. 
The residuals  smoothly
change in one hour, and  an event at $\sim$9 km/s develops.
Its deepest observed intensity of 0.068 depth is achieved during the last hour. A
broad red wing seems to develop during the four observing hours.

\begin{figure} 
\centering
\scalebox{0.345}{\includegraphics[angle=-90]{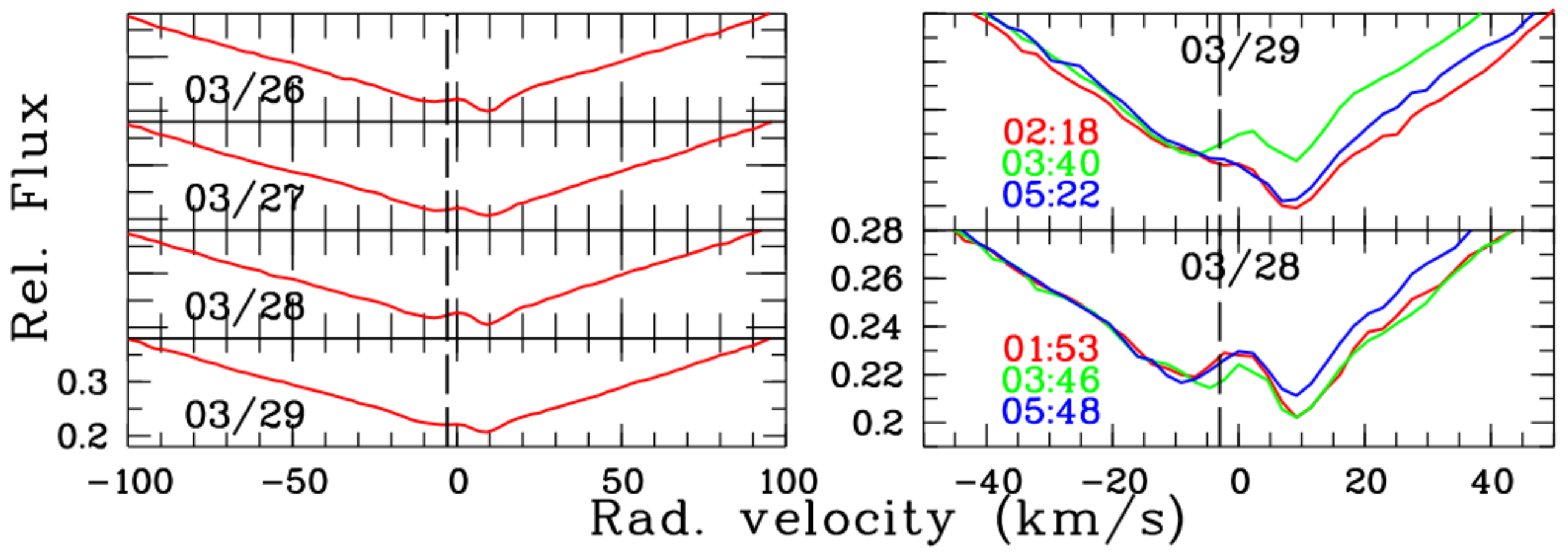}}
\caption{Ca {\sc ii} K spectra obtained with the MPG/ESO 2.2 m telescope
  from  March  26   to  March  29.  Right:  average   of  the  nightly
  spectra. Left: individual  spectra taken during the  nights March 28
  and  March  29.  Colors  correspond  to  the  spectra  obtained  at
  different times (UT are shown).}
\label{lasilla} 
\end{figure}

\noindent {\em MPG/ESO 2016 March.}  The  Ca {\sc ii} K line exhibited
a  pronounced variation  from  night to  night  and within  individual
nights on timescales shorter than two hours. No regular pattern
can be  inferred from  the variations.   Figure \ref{lasilla}  shows the
average of the  spectra taken during each of the  four nights. All nights
show a distinct redshifted event  with respect to the template spectrum. Furthermore,
the  strength of  the events  changes strikingly  from one  spectrum to
another, with time differences as short  as 90 minutes. To exemplify this
behavior, Fig.  \ref{lasilla} shows the individual  spectra from March
28 and March  29. Events at $\sim$9 km/s and depths  of 0.118 (March
28)  and  0.093  (March  29)  are detected.  There  might  exist  some
blueshifted absorptions, at  least on March 28,  which vary in velocity,
-4 to -9 km/s, and tiny changes in the depth of 0.04--0.05.

\section{Discussion}

The  variable  absorptions  at  10-20 km/s  from  the  star,  with velocity
dispersions of $\sim$10-20 km/s,  depths of $\sim$0.05-0.1, and equivalent
widths of $\sim$20 m\AA, that were observed in  $\phi$ Leo behave similarly  to those  attributed  to  exocomets  in $\beta$  Pic.  Nonetheless,  it  is
reasonable to  consider other alternatives because  a remarkable difference
between  the two stars  is that $\phi$ Leo is not associated with a massive 
debris disk. This  
might be  due to the age of $\phi$ Leo
because  warm  debris  disks  decline   rapidly  for  stars  older  than  a
characteristic time of $\sim$150 Myr \citep{rieke05}. In addition, as noted
before, it is unknown whether the star is surrounded by a cold debris disk,
which in  any  case  declines   rapidly  for  stars  older  than  a
characteristic  time  of  $\sim$400 Myr  \citep{su06}.  Furthermore,  some
relatively old stars with exocomet signatures do not possess a debris
disk \citep{roberge08}.

For $\beta$ Pic, phenomena involving the stellar atmosphere
as    the    origin    of    the    events    were    soon    excluded
\citep{lagrangehenri90}.  This  seems also to  be the case  of $\phi$
Leo,  since we  would expect  similar behavior  in other  photospheric
lines, which is not the case. Moreover, given the very short timescales
of  the  variations,  an  origin  in the  intersellar  medium  is  not
plausible.   Mass-loss rates  resulting from  weak radiatively  driven
winds in A  stars are not expected to be  significant around late-type
A7IV  stars such as  $\phi$  Leo.   A wind  origin  like this
was excluded  for
$\beta$  Pic  \citep{lagrangehenri92}, although  it  has  not been  for
events in other stars  \citep{redfield07}.  Sporadic mass-loss events,
that is,  clumps generated in a  hypothetical stellar wind  and rotating
with  the star,  or  moving  outward with  the  wind, are expected
to  produce
emission lines moving from blue to  red across the center of the line,
or be present  in the blue and red wings of  the wind lines. This
is  typically   observed  in  relatively   hot  stars,  like   Be  stars
\citep{porter03}.  However,  this behavior is not  observed in $\phi$
Leo.

The ratio  of the event  depths observed  in the Ca  {\sc ii} K  and H
lines,  $\sim$1-1.5, differs  from  the oscillator  strength ratio  of
these lines (=2), suggesting that the variable absorptions arise
at least partly from optically thick  clumpy gas that covers a small fraction
of the  stellar surface,  again similar  to $\beta$  Pic or  HD 172555
\citep{lecavelier97,kiefer14}. These  results, together with  the high-velocity  broad  line  wings  suggested  in  some  spectra,  makes  it
plausible that the $\phi$ Leo events trace the evaporation of
comet-like bodies in  the CS environment. In this  scenario, the broad
wings could  be produced by  a sort of  comet-like coma or  by several
unresolved events.
 
Assuming a  cometary origin, the  similar dynamical properties  of the
events could  point to  a disruption  of a  larger primordial  body by
tidal  forces in  a  near-stellar encounter,  or to  a  family of  comets
\citep{beust96}  driven into  the vicinity  of the  star by  a planet.
Several   not  mutually  exclusive mechanisms might   
deliver the  bodies: Kozai-Lidov \citep{Naoz2016},  secular resonances
\citep{Levison1994}, mean motion  resonances \citep{Bm1996}, or direct
scattering by an eccentric planet  \citep{beust91}.   We note that
  only a  future improvement of  the event statistics in  $\phi$ Leo
  will  help  to  better  constrain   these   dynamical
  mechanisms. The orbit  of the infalling bodies can  only be roughly
estimated  since  the  radial  velocity  is  degenerated  between  the
pericenter distance and the angle formed between the axis of the orbit
and the  line of  sight (true anomaly).  Thus, the  largest pericenter
distances at which the  bodies can originate (for a  3.25 M$_\odot$ star,
estimated  from the  stellar  parameters in  Sect. \ref{hr4368})  is
$\sim$  20-70 au.   The observed  infall velocities,  $\sim$0-20 km/s,
possibly with  some wings of up  to 200 km/s, are  much lower that  the free-fall velocity at reasonable pericenter distances, 79 km/s and 177 km/s
at 50 and 10 $R_\star$, respectively. The most plausible
option apparently is that the presumable exocomets follow parabolic to hyperbolic
orbits that cross  the line  of sight, in which  case the  main absorption event
 and   the  broad  wings   at  larger  velocities   can  be
explained.   The crossing  distances must be  close enough  to the
  star   to  allow   refractory  material   to  sublimate.   Following
  \cite{beust98}, the shortest crossing distance  would be $\sim$0.21 au using a
  FWHM of 20 km/s and assuming that the  source of broadening 
the line is Keplerian. Alternatively, if we assume thermal equilibrium 
\citep{beust96}, the distance 
at which dust sublimates, taking as a typical value $T_{\rm sub} = 1500$ K, would 
be $\sim$0.46 au or $\sim$0.33 au for albedos 0.0 and 0.5, respectively. These values depend on the composition of the grains.

With respect to the stable
Ca {\sc  ii}  component, we can exclude it originates in the  local interstellar medium since  interstellar lines are
considerably narrower, the line of sight traverses the Leo cloud, which
has a  centroid radial velocity  of 1.75 km/s  \citep{redfield08}, and
its equivalent  width gives a  lower limit  for the column  density of
$N(H) \gtrsim$ 10$^{20} {\rm cm}^{-2}$,  much higher than the expected
interstellar column densities in the Leo direction \citep{redfield00}.
This means that the component might originate in one of the scenarios suggested for
the stable  CS gas disk in  $\beta$ Pic: stellar  winds, star-grazing
comet evaporation, or  grain evaporation near the star or  in the disk
\citep[see][]{fernandez06}, although we  note that
the stable $\phi$ Leo component is  broader and not as sharp as in
$\beta$ Pic  \citep{lagrangehenri92}.  On the other  hand, practically
all stars showing a triangular shape rotate at very high velocities,
$v \sin i \gtrsim $ 200 km/s. At these velocities the structure of the
rotating stars and  the inclination angle with respect to  the line of
sight   affect    their   location   on    color-magnitude   diagrams
\citep{bastian09}, and the  induced oblateness of the  star produces a
gravity  darkening that results  from the  temperature gradient  from the
stellar  equator to  the poles.  A preliminary test  for which
Kurucz  models were combined with  a temperature  gradient from  the equator  to the
poles indicated that  the strength of the  additional triangular absorption
decreases.  A  deeper analysis 
has  been undertaken  (Montesinos et
al., in preparation).

\section{Conclusions}
Our 
intensive 
monitoring of $\phi$  Leo 
showed  that its spectrum is very  rich in  redshifted absorption  events, which
might  be  accompanied   in  some  cases  by  broad   wings  and  even
blueshifted  absorptions. These  
sporadic 
events  are similar  to those in $\beta$  Pic and can be most plausibly explained   as    exocomets that graze the star  or  fall  onto  the
stellar surface. Assuming  this scenario, it  is intriguing how 
a  relatively old
 500-900 Myr star such  as $\phi$  Leo, which does not have  any known  associated  debris disk,  can
possess  such  a  rich  environment  that hosts  minor  bodies.   Another
interesting aspect  is   the  origin  of  what might be   a
triangular-shaped stable  CS absorption  component in the  Ca {\sc
  ii}  lines.  Additional   monitoring  is clearly needed  to  better
characterize the  sporadic events and the stable  component by
comparing them with similar stars.

\begin{acknowledgements}
Based on  observations made with  the Mercator Telescope,  operated on
the  island of  La Palma  by the  Flemmish Community,  at the  Spanish
Observatorio  del  Roque   de  los  Muchachos  of   the  Instituto  de
Astrofísica de Canarias. 
H.C., C.E., G. M., B. M., I. R., and E.V. are
supported by Spanish grant AYA 2014-55840-P. J.O. acknowledges support from ALMA/Conicyt Project 31130027. 
O.A.\ is F.R.S.-FNRS Research Associate. We thank the referee H. Beust for his
constructive comments.
\end{acknowledgements}

%
\bibliographystyle{aa} 
\bibliography{hr4368} 

\end{document}